%% file: Main.tex
\documentclass[10pt,journal,letterpaper]{IEEEtran}

\usepackage{epstopdf}
\usepackage{graphicx}
\usepackage{stmaryrd}
\usepackage{amsmath,amsthm,amssymb}
\usepackage{multirow,url}
\usepackage{color,cite}
\usepackage{algorithm}
\usepackage{algorithmicx}
\usepackage{algpseudocode}
\usepackage{setspace}

\newtheorem{lem}{Lemma}

\newtheorem{thm}{Theorem}
\newtheorem{defn}{Definition}
\newtheorem{cor}{Corollary}

\ifodd 0
\newcommand{\com}[1]{\textbf{\color{red} (COMMENT: #1)}} %comment of the text
\newcommand{\comg}[1]{\textbf{\color{green} (COMMENT: #1)}}
\newcommand{\response}[1]{\textbf{\color{magenta} (RESPONSE: #1)}} %response to comment
\else

\newcommand{\com}[1]{}
\newcommand{\comg}[1]{}
\newcommand{\response}[1]{}
\fi
\begin{document}

\title{Imitation-based Social Spectrum Sharing}

\author{Xu Chen, \emph{Member, IEEE}, and Jianwei Huang, \emph{Senior Member, IEEE} \thanks{Xu Chen is with the School of Electrical, Computer and Energy Engineering, Arizona State University, Tempe, Arizona, USA (email:xchen179@asu.edu). The work was mainly done when he was with the Chinese University of Hong Kong.

Jianwei Huang is with the Network Communications and Economics Lab, Department of Information Engineering, the Chinese University of Hong Kong (email:jwhuang@ie.cuhk.edu.hk).

Part of the results have appeared in IEEE WiOpt conference \cite{chen2012imitative}.
}

}

%\IEEEcompsoctitleabstractindextext{
%
%
%\begin{IEEEkeywords}
%Cognitive radio, distributed spectrum access, social intelligence, imitation, imitation equilibrium.
%\end{IEEEkeywords}
%%\IEEEpeerreviewmaketitle
%}

\maketitle

\thispagestyle{empty}
\allowdisplaybreaks
\pagestyle{empty}

%\begin{spacing}{1.58}

\input{Abstract}
\input{Introduction}

\input{System}
\input{Imitation}

\input{Convergence2}

\input{Innovation}

\input{Results}
\input{Conclusion}

\bibliographystyle{ieeetran}
\bibliography{DynamicSpectrum}

\input{Appendix}

\end{document}

%% file: Abstract.tex
%!TEX root = main.tex
%SourceDoc main.tex

\begin{abstract}
Dynamic spectrum sharing is a promising technology for improving the spectrum
utilization. In this paper, we study how secondary users can share the spectrum in a distributed fashion based on social imitations. The imitation-based mechanism leverages the social intelligence of the secondary user crowd and only requires a low computational power for each individual user.  We introduce the information sharing graph to model the social information sharing relationship among the secondary users. We propose an imitative spectrum access mechanism on a general information sharing graph such that  each secondary user first estimates its expected throughput based on local observations, and then imitates the channel selection of another neighboring user who achieves a higher throughput. We show that the imitative spectrum access mechanism converges to an imitation equilibrium, where no beneficial imitation can be further carried
out on the time average. Numerical results show that  the imitative spectrum access mechanism can achieve efficient spectrum utilization and meanwhile provide good fairness across secondary users.
\end{abstract}

%% file: Introduction.tex
%!TEX root = main.tex
%SourceDoc main.tex

\section{Introduction}

Dynamic spectrum sharing is envisioned as a promising technique to alleviate
the problem of spectrum under-utilization  \cite{key-1}. It enables unlicensed
wireless users (secondary users) to opportunistically access the licensed
channels owned by legacy spectrum holders (primary users), and thus
can significantly improve the spectrum efficiency \cite{partridge2011realizing}.

A key challenge of dynamic spectrum sharing is how to share
the spectrum resources in an intelligent fashion. Most of existing works in dynamic spectrum sharing networks focus on exploring the \emph{individual intelligence} of the secondary users. A common modeling approach is to consider that secondary users are \emph{fully rational}, and model their interactions as noncooperative
games (e.g., \cite{key-21,key-4,li2010market,chen2012spatial,niyato2009dynamics2} and many others).
Nie and Comniciu in \cite{key-21} designed a self-enforcing distributed
spectrum access mechanism based on potential games. Niyato and Hossain
in \cite{key-4,niyato2009dynamics} studied a price-based spectrum access mechanism for
competitive secondary users. Chen and Huang in \cite{chen2012spatial} proposed a spatial spectrum access game framework for distributed spectrum sharing problem with spatial reuse.  Li \emph{et al.} in \cite{li2010market} proposed a game theoretic framework to achieve incentive compatible multi-band sharing among the secondary users.

When not knowing spectrum information such as channel availability, secondary users need to learn the network environment and adapt the spectrum access decisions accordingly. Han \emph{et al.} in \cite{key-25} used no-regret learning to solve this problem, assuming that the users' channel selections are common information. When users' channel selections are not observable, authors in \cite{key-27,key-29} designed multi-agent multi-armed bandit learning algorithms to minimize the expected performance loss of distributed spectrum access

The common assumption of all the above work
is that secondary users act fully rationally based on individual intelligence. To have full rationality, a
user typically needs to have a high computational power to collect
and analyze the network information in order to predict
other users' behaviors. This is often not feasible due to the limitations of today's wireless devices.

%Another body of related work focused on the design of
%spectrum access mechanisms assuming \emph{bounded rationality}
%of secondary users, i.e., each user tries to improve
%its strategy adaptively over time. Authors in \cite{ikey-25}, \cite{key-26} used
%no-regret learning to solve this problem, assuming that
%the users' channel selections are common information. Authors in \cite{li2010customer } and \cite{chen2011adaptive} designed adaptive channel access mechanisms based on users' real-time spectrum access experiences.  When users'
%channel selections are not observable, authors in \cite{ikey-28,ikey-29} designed multi-agent multi-armed bandit learning
%algorithm to minimize the expected performance loss of
%distributed spectrum access.

In this paper, we will explore the \emph{social intelligence} of the secondary users based on social interactions.  The motivation is to overcome the limited capability of today's wireless devices by leveraging the wisdom of secondary user crowds.  In fact, the emergence of social intelligence has been observed in many social interactions of animals \cite{sumpter2010collective} and has been utilized for engineering algorithm design. For example, Kennedy and Eberhart intended the particle swarm optimization  algorithm by simulating social movement behaviors in a bird flock \cite{kennedy1995particle}. Pham \emph{et al.} developed the bees algorithm by mimicing the food foraging behaviors of honey bees \cite{pham2006bees}. The understanding of human social phenomenon also sheds new light into the design of more efficient engineering systems such as wireless communication networks. For example, Wang and Yoneki in \cite{wang2011impact} exploited the social structure such as centrality and community to design efficient forwarding algorithms in opportunistic networks. The trust mechanism in human social activities has also been utilized to promote node cooperation in mobile ad hoc networks \cite{cho2010survey}.

In this paper, we will design distributed spectrum access mechanism based on \emph{imitation}, which is also a common  phenomenon in many social animal and human interactions \cite{wyrwicka1996imitation}. Imitation is simple (just follow a successful action  of another user) and turns out to be an efficient strategy
in many applications \cite{SCHLAG2009}. For example, Schlag in \cite{key-IM5} used
imitation to solve the multi-armed bandit problem. Lopes \emph{et al.} in \cite{key-IM8} designed an efficient imitation-based social learning
mechanism for robots. Alos-Ferrer and Weidenholzer in \cite{alos2008contagion} investigated the imitation strategy for the network coordination game.  Imitation in wireless networks, however, has several fundamental differences from the previous approaches. For example, when too many wireless users imitate the same channel choice, then severe congestion in spectrum utilization will occur. Furthermore, different wireless users may experience different channel conditions due to the local environmental factors such as fading. Hence it is challenging to design an efficient mechanism based on imitation to deal with the user heterogeneity.

Recently, Iellamo \emph{et al.}~in \cite{key-IM6} proposed an imitation-based
spectrum access mechanism for spectrum sharing networks, by assuming that all the secondary users are homogeneous (i.e., they experience the same channel condition) and the true expected throughput on a channel is known by a secondary user once the user has chosen the channel. In this paper, we relax these restrictive assumptions
and design an imitative spectrum access mechanism based on user's
\emph{local observations} such as the realized data rates and transmission collisions.  The key idea is that each user applies the
maximum likelihood estimation to estimate its expected throughput,
and imitates another neighboring user's channel selection if neighbor's estimated
throughput is higher. Moreover, as imitation requires limited information sharing, we introduce the information sharing graph to model the social information sharing relationship among the secondary users. For example, in practical wireless systems it is often the case that a user can only receive message broadcasting from a subset of users that are close enough due to the geographical constraint. Moreover, we also generalize the proposed imitation based spectrum access mechanism to the case that secondary users are heterogeneous.  The main results and contributions of this paper
are as follows:
\begin{itemize}
\item \emph{Imitative Spectrum Access}: We propose a novel imitation-based
distributed spectrum access mechanism on a general information sharing graph. Each secondary user first estimates its expected throughput based on local observations, and then chooses to imitate a better neighbor. The imitation-based mechanism leverages the social intelligence of the secondary user crowd and only requires a low computational power for each individual user.

\item \emph{Convergence to the Imitation Equilibrium}: We show that the imitative
spectrum access mechanism converges to the imitation equilibrium,  wherein
no imitation can be further carried out on the time average. When the information sharing graph is connected, we show that the imitation equilibrium corresponds to a fair channel allocation, such that all the users achieve the same throughput in the asymptotic case.

\item \emph{Imitative Spectrum Access with User Heterogeneity}: We
further design an imitation-based spectrum access mechanism
with user heterogeneity, where different users
achieve different data rates on the same channel. Numerical results show that the proposed mechanism achieves up-to $530\%$  fairness improvement with at most $20\%$ performance loss, compared with the centralized optimal solution. This demonstrates that the proposed imitation-based mechanism can achieve efficient spectrum utilization and meanwhile provide good fairness across secondary users.
\end{itemize}

The rest of the paper is organized as follows. We introduce the system
model in Section \ref{sec:System-Model}. We then present the imitative spectrum access mechanism in Section \ref{sec:Imitative-Spectrum-Access}, and study the dynamics and convergence of the imitative spectrum access mechanism in Section \ref{convergence2}. We proposed imitative spectrum access mechanism with user heterogeneity, and illustrate
the performance of the proposed mechanisms through numerical results
in Sections \ref{ISA2} and \ref{sec:Simulation-Results}, respectively, Finally, we conclude the paper in Section \ref{sec:Conclusion}.

%% file: System.tex
%!TEX root = main.tex
%SourceDoc main.tex

\section{\label{sec:System-Model}System Model}
In this part, we first discuss the system model of distributed spectrum sharing, and then introduce the information sharing graph for the imitation mechanism.

%\begin{figure}[tt]
%\centering
%\includegraphics[scale=1.4]{TwoNets}
%\caption{\label{fig:TwoNets} Illustration of imitation-based spectrum sharing system, which consists of spectrum sharing network and information sharing graph. The users (representing by red dots) in both networks are the same. In the spectrum sharing network, two users can generate inteference with each other if they are connected by a black edge. In the information sharing graph, two users can exchange information with each other if they are connected by a green edge. }
%
%\end{figure}

\subsection{\label{sec:System-Model1}Spectrum Sharing System Model}
\begin{figure}[tt]
\centering
\includegraphics[scale=0.45]{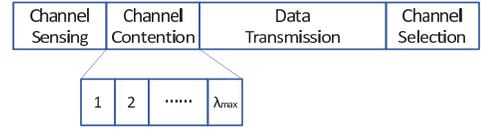}
\caption{\label{fig:Time-slot-structure} Multiple stages in a single time slot.}
\end{figure}

%Wireless spectrums are often licensed to service providers based on long-term agreements. A service provider will serve its primary users using its licensed spectrum. Due to the stochastic nature of primary users' traffic, the licensed spectrum may not be fully utilized at all locations and all times. Field measurements by Shared Spectrum Cooperation in Chicago area shows that the overall average utilization of a wide range of different types of spectrum bands is lower than $20\%$ \cite{SOM2005}. Based on the recent policy reforms in several countries (such as the FCC's ruling for the TV white space \cite{FCC}), secondary unlicensed users equipped with cognitive radios can opportunistically access the tentatively unused channels in order to improve the  overal spectrum utilization.
We consider a spectrum sharing network with a set $\mathcal{M}=\{1,2,...,M\}$
of independent and\emph{ stochastically heterogeneous} licensed channels.
A set $\mathcal{N}=\{1,2,...,N\}$ of secondary users
try to opportunistically access these channels, when the channels
are not occupied by primary (licensed) transmissions. For simplicity, we assume that all secondary users accessing the same channel will interfere with each other (i.e., the interference graph under the protocol interference model \cite{gupta2000capacity} is fully meshed). The case with the spatial reuse (i.e., the interference graph can be partially meshed) will be considered in a future work. The system model
has a slotted transmission structure as in Figure \ref{fig:Time-slot-structure} and is described as follows.

1) \emph{Channel State}: the channel state for a channel $m$ during a
time slot $\tau$ is
\[S_{m}(\tau)=
\begin{cases}
0, & \mbox{if channel \ensuremath{m} is occupied}\\
 & \mbox{by primary transmissions,}\\
1, & \mbox{if channel \ensuremath{m} is idle.}\end{cases}\]

2) \emph{Channel State Changing}: for a channel $m$, we assume that
the channel state is an i.i.d. Bernoulli random variable, with an
idle probability $\theta_{m}\in(0,1)$ and a busy probability $1-\theta_{m}$.
This model can be a good approximation of the reality if the time slots
for secondary transmissions are sufficiently long or the primary transmissions
are highly bursty \cite{key-27}. The motivation of considering the i.i.d. channel state model is to focus our analysis on the spectrum contention due to secondary users' dynamic channel selections. However, numerical results show that the proposed mechanism also works well in the Markovian channel environment where channel states have correlations between time slots. Please refer to Section \ref{MCE} for a detailed discussion.

3) \emph{Heterogeneous Channel Throughput}: if a channel $m$ is idle, the achievable data rate by a secondary user in each time slot $b_{m}(\tau)$ evolves according
to an i.i.d. random process with a mean $B_{m}$, due to the local
environmental effects such as fading \cite{rappaport1996wireless}. For example, we can compute the data rate
$b_{m}(\tau)$ according to the Shannon capacity as\begin{equation}
b_{m}(\tau)=E_{m}\log_{2}\left(1+\frac{\eta_{n}h_{m}(\tau)}{\upsilon_{m}}\right),\label{eq:dd}\end{equation}
where $E_{m}$ is the bandwidth of channel $m$, $\eta_{n}$ is
the fixed transmission power adopted by user $n$ according to the requirements such as the primary user protection, $\upsilon_{m}$ denotes
the background noise power, and $h_{m}(\tau)$ is the channel gain. In a Rayleigh fading
channel environment, the channel gain $h_{m}(\tau)$ is a random variable that follows the exponential distribution \cite{rappaport1996wireless}. Here we first consider the homogeneous user case that all users achieve the same mean data rate on the same channel (but users can achieve different data rates on different channels). In Section \ref{ISA2}, we will further
consider the heterogeneous user case that different users can achieve different mean data rates even on the same channel. This will allow
users to have different transmission technologies, choose
different coding/modulation schemes, and experience different
channel conditions.

4) \emph{Time Slot Structure}: each secondary user $n$ executes the
following stages synchronously during each time slot:
\begin{itemize}
\item \emph{Channel Sensing}: sense one of the channels based on the channel
selection decision generated at the end of previous time slot (see below). Access the channel if it is idle.
\item \emph{Channel Contention}: use a backoff mechanism to resolve collisions when
multiple secondary users access the same idle channel\footnote{For ease of exposition, we adopt the backoff mechanism as an example. Our analysis can apply to many other medium access control (MAC) schemes such as TDMA.}. The contention stage of a time slot is divided into $\lambda_{\max}$
mini-slots\footnote{Note that in general the length of a mini-slot is much smaller than the length of spectrum sensing and access period in a time slot. For example, for IEEE 802.11af systems (also known as WhiteFi Networks), the length of a mini-slot is $4$ microseconds and the spectrum sensing duration is $0.5$ milliseconds \cite{flores2013ieee}.} (see Figure~\ref{fig:Time-slot-structure}), and
user $n$ executes the following two steps.
\emph{First}, count down according to a randomly and uniformly chosen integral backoff
time (number of mini-slots) $\lambda_{n}$ between $1$ and $\lambda_{\max}$. \emph{Second,} once the timer expires, transmit RTS/CTS messages if the channel is clear (i.e., no ongoing transmission).
Note that if multiple users choose the same backoff value $\lambda_{n}$, a collision
will occur with RTS/CTS transmissions and no users win the channel contention.
\item \emph{Data Transmission}:  transmit data packets if the RTS/CTS message exchange is successful (i.e., the user wins the channel contention).
\item \emph{Channel Selection}: choose a channel to access in the next time slot
according to the imitative spectrum access mechanism (introduced in
Section \ref{sec:Imitative-Spectrum-Access}).
\end{itemize}
%\end{itemize}

Suppose that $k_{m}$ users choose the same idle channel $m$ to access. Then
the probability that a user $n$ (out of the $k_{m}$ users) successfully grabs the
channel $m$ is\begin{eqnarray*}
g(k_{m}) & = & Pr\{\lambda_{n}<\min_{i\neq n}\{\lambda_{i}\}\}\\
 & = & \sum_{\lambda=1}^{\lambda_{\max}}Pr\{\lambda_{n}=\lambda\}Pr\{\lambda<\min_{i\neq n}\{\lambda_{i}\}|\lambda_{n}=\lambda\}\\
 & = & \sum_{\lambda=1}^{\lambda_{\max}}\frac{1}{\lambda_{\max}}\left(\frac{\lambda_{\max}-\lambda}{\lambda_{\max}}\right)^{k_{m}-1},\end{eqnarray*}
which is a decreasing function of the total contending users
$k_{m}$. Then the long-run expected throughput of a secondary user $n$ choosing a
channel $m$ is given as\begin{equation}
U_{n}=\theta_{m}B_{m}g(k_{m}).\label{eq:u1}\end{equation}

\subsection{\label{ISG}Social Information Sharing Graph}

In order to carry out imitations, we assume that there exists a common control channel for the information exchange among secondary users\footnote{There are several approaches for establishing a common control channel in cognitive radio networks, e.g., sequence-based rendezvous \cite{dasilva2008sequence}, adaptive channel hopping \cite{bian2011control} and user grouping \cite{lazos2009spectrum}. Please refer to \cite{lo2011survey} for a comprehensive survey on the research of common control channel establishment in cognitive radio networks.}. As an alternative, we can adopt the proximity-based communication approach \cite{drost2004proximity}, such that secondary users equipped with the radio interfaces such as near field communication (NFC)/bluetooth/WiFi Direct can communicate with each other directly for information exchange. Since information exchange typically would incur an overhead such as the extra energy consumption, it is important to design a proper incentive mechanism for stimulating collaborative information exchange among secondary users. One possible approach is to design mechanisms such that users receive exogenous incentives for cooperation. For example, a payment based incentive mechanism \cite{anderegg2003ad} compensates users' contributions by rewarding them with virtual currency. In reputation based incentive mechanisms \cite{michiardi2002core}, users' cooperative behaviors are monitored by some centralized authority or collectively by the whole user population, so that any user's selfish behaviors would be detected and punished. In general, such an approach requires centralized infrastructures (e.g., secondary base-station/access point), which would incur a high system overhead and may not be feasible in our context of distributed spectrum sharing.

The centralized infrastructures are not available, motivated by the observation that the hand-held devices are typically carried by human beings, we can leverage the endogenous incentive which comes from the intrinsic social relationships among users to promote effective and trustworthy cooperation. For example, when a user
is at home or work, typically family members, neighbors, colleagues, or friends are nearby. The user can then
exploit the social trust from these neighboring users to achieve effective cooperation for information exchange. Indeed, with the explosive growth of online social networks such as Facebook and Twitter, more and more people are actively involved in online social interactions, and social connections among people are being extensively broadened. This has opened up a new avenue to integrate the social interactions for cooperative networking design.

Specifically, we introduce the social information sharing graph $\mathcal{G}=\{\mathcal{N},\mathcal{E}\}$
to model cooperative information exchange relationships due to the social ties among the secondary users. Here the vertex set is the same as the user
set $\mathcal{N}$, and the edge set is given as $\mathcal{E}=\{(n,m):e_{nm}=1,\forall n,m\in\mathcal{N}\}$
where $e_{nm}=1$ if and only if users $n$ and $m$ have social
tie between each other, e.g., kinship, friendship, or colleague relationships. Furthermore, for a pair of users $n$ and $m$ who have a social
edge between them on the social graph, we formalize the strength of
social tie as $\delta_{nm}\in[0,1]$, with a higher value of $\delta_{nm}$ being
a stronger social tie. Each secondary user $n$ can specify a cooperation threshold $\varphi_n$ and is willing to share information with those users with whom he has a high enough social tie above the cooperation threshold $\varphi_n$. Moreover, to thwart the potential attacks of releasing false channel information by malicious users and enhance the security level of imitation based spectrum access, each secondary user $n$ can set a trust threshold $\eta_{n}$ and choose to trust the information from those users having a high enough social tie above the trust threshold $\eta_{n}$. In the sequel, we denote the neighborhood of user $n$ for effective and trustworthy information sharing as $\mathcal{N}_{n}\triangleq\{k:e_{nk}=1\mbox{ and }\delta_{nk}\geq\eta_{n}\mbox{ and }\delta_{kn}\geq\varphi_{k}\}$. In terms of  implementation, the social relationship identification procedure can be carried out prior to the imitative  spectrum access. Specifically, two secondary users can locally initiate the ``matching'' process to detect the common social features between  them. For example, two users can match their contact lists. If they have the phone numbers of each other or many of their phone numbers are the same, then it is very likely that they know each other. As another example, two device users can match their home and working addresses and identify whether they are neighbors or colleagues. To preserve the privacy of the secondary users, the private set intersection and homomorphic encryption techniques proposed in \cite{zhang2012fine,von2008veneta} can be adopted to design a privacy-preserving social relationship identification mechanism.

We should emphasize that when it is difficult to leverage the social trust among some secondary users and the centralized infrastructures are not available, similar to the file sharing in the P2P systems \cite{cohen2003incentives}, we can adopt the Tit-for-Tat mechanism for information sharing. Specifically, based on the principle of reciprocity, a secondary user will always share information with its partner as long as its partner (i.e., another user) also shares information with it. If the partner refuses to share information, the user will punish its partner by not sharing information either. As a result, the partner would suffer and learn to share information with the user again.  Notice that since the imitative spectrum access mechanism can work on a generic social information sharing graph, we can also use a hybrid approach of several different schemes mentioned above for establishing the information sharing relationships among the secondary users.

Since our analysis is from secondary users' perspective, we will use  terms ``secondary user'' and ``user'' interchangeably in the following sections.

%% file: Imitation.tex
%!TEX root = main.tex
%SourceDoc main.tex

\section{\label{sec:Imitative-Spectrum-Access}Imitative Spectrum Access Mechanism}

We now apply the idea of imitation to design an efficient distributed
spectrum access mechanism, which utilizes user's local estimation
of its expected throughput. Each user randomly chooses a neighboring user in the information sharing graph, and follows the neighbor's channel selection if the neighbor's throughout is better than its.

\subsection{\label{MLE}Expected Throughput Estimation}
\begin{figure}[tt]
\centering
\includegraphics[scale=0.45]{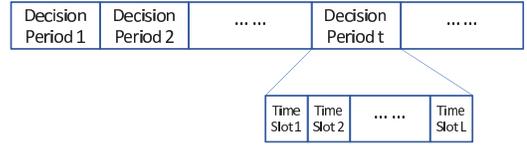}
\caption{\label{fig:time}The period structure of maximum likelihood estimation of various system parameters.}
\end{figure}

In order to imitate a successful action, a user needs to compare its and other users' performances (throughputs). In
practice, many wireless devices only have a limited view of the network environment due to hardware constraints. To
incorporate the effect of incomplete network information, we first introduce the maximum likelihood
estimation (MLE) approach to estimate user's expected throughput based on its local observations.
We choose MLE mainly due to the efficiency and the ease of implementation of this method \cite{key-IM1}. To achieve accurate local estimation based on local observations, a user
needs to gather a large number of observation samples. This motivates
us to divide the spectrum access time into a sequence of \emph{decision
periods }indexed by $t(=1,2,...)$, where each decision period consists
of $L$ time slots (see Figure \ref{fig:time} for an illustration). During a single decision period, a user accesses
the \emph{same} channel in all $L$ time slots. Thus the total number
of users accessing each channel does not change within a decision
period, which allows users to learn the environment.

According to (\ref{eq:u1}), a user's expected throughput during decision
period $t$ depends on the probability of grabbing the channel $g(k_{m}(t))$
on that period, the channel idle probability $\theta_{m}$, and the
mean data rate $B_{m}$.

\subsubsection{MLE of Channel Grabbing Probability $g(k_{m}(t))$}

At the beginning of each time slot $\tau(=1,...,L)$ of a decision period
$t$, we assume that a user $n$ chooses to sense the same channel $m$. If the channel
is idle, the user will contend to grab the channel according to the
backoff mechanism in Section \ref{sec:System-Model}. At the end of
each time slot $\tau$, a user observes $S_{n}(t,\tau)$, $I_{n}(t,\tau)$,
and $b_{n}(t,\tau)$. Here $S_{n}(t,\tau)$ denotes the state of the chosen
channel (i.e., whether occupied by the primary traffic), $I_{n}(t,\tau)$
indicates whether the user has successfully grabbed the channel, i.e.,\[
I_{n}(t,\tau)=\begin{cases}
1, & \mbox{if user $n$ successfully grabs the channel}\\
0, & \mbox{otherwise,}\end{cases}\]
and $b_{n}(t,\tau)$ is the received data rate on the chosen channel
by user $n$ at time slot $\tau$. Note that if $S_{n}(t,\tau)=0$ (i.e., the channel is occupied by the
primary traffic), we set $I_{n}(t,\tau)$ and $b_{n}(t,\tau)$ to be
$0$. At the end of each decision period
$t$, each user $n$ will have a set of local observations $\Omega_{n}(t)=\{S_{n}(t,\tau),I_{n}(t,\tau),b_{n}(t,\tau)\}_{\tau=1}^{L}$.

When channel $m$ is idle (i.e., no primary traffic), consider $k_{m}(t)$
users contend for the channel according to the backoff mechanism in
Section \ref{sec:System-Model}. Then a particular user $n$ out of these $k_{m}(t)$
users grabs the channel with the probability $g(k_{m}(t))$. Since there are
a total of $\sum_{\tau=1}^{L}S_{n}(t,\tau)$ rounds of channel contentions
in the period $t$ and each round is independent, the total number
of successful channel captures $\sum_{\tau=1}^{L}I_{n}(t,\tau)$ by user $n$ follows
the Binomial distribution. User $n$ then computes the likelihood
of $g(k_{m}(t))$, i.e., the probability of the realized observations
$\Omega_{n}(t)$ given the parameter $g(k(t))$ as
\begin{multline}
\mathcal{L}[\Omega_{n}(t)|g(k_{m}(t))] =  \left(\begin{array}{c}
\sum_{l=1}^{L}S_{n}(t,l)\\
\sum_{l=1}^{L}I_{n}(t,l)\end{array}\right)g(k_{m}(t))^{\sum_{l=1}^{L}I_{n}(t,l)}\nonumber \\
 \times(1-g(k_{m}(t)))^{\sum_{l=1}^{L}S_{n}(t,l)-\sum_{l=1}^{L}I_{n}(t,l)}.\label{eq:MLE1}
\end{multline}

Then MLE of $g(k_{m}(t))$ can be computed by maximizing the log-likelihood
function $\ln\mathcal{L}[\Omega_{n}(t)|g(k_{m}(t))]$, i.e., $\max_{g(k_{m}(t))}\ln\mathcal{L}[\Omega_{n}(t)|g(k_{m}(t))]$.
By the first order condition, we obtain the optimal solution as $\tilde{g}(k_{m}(t))={\sum_{\tau=1}^{L}I_{n}(t,\tau)}/{\sum_{\tau=1}^{L}S_{n}(t,\tau)},$
which is the sample averaging estimation. When the length of decision
period $L$ is large, by the central limit theorem, we know that $\tilde{g}(k_{m}(t))\sim\mathcal{N}\left(g(k_{m}(t)),\frac{g(k_{m}(t))(1-g(k_{m}(t)))}{\sum_{\tau=1}^{L}S_{n}(t,\tau)}\right),$
where $\mathcal{N}(\cdot)$ denotes the normal distribution.

\subsubsection{MLE of Channel Idle Probability $\theta_{m}$}

We next apply the MLE to estimate the channel idle probability $\theta_{m}.$
Since the channel state $S_{n}(t,\tau)$ is i.i.d over different time
slots and different decision periods, we can improve the estimation
by averaging not only over multiple time slots but also over multiple periods.

Similarly with MLE of $g(k_{m}(t))$, we first compute one-period MLE
of $\theta_{m}$ as $\hat{\theta}_{m}=\frac{\sum_{\tau=1}^{L}S_{n}(t,\tau)}{L}$. When the length of decision period $L$ is large, we have that $\hat{\theta}_{m}$ follows the normal distribution with the mean $\theta_{m}$, i.e., $\hat{\theta}_{m}\sim\mathcal{N}\left(\theta_{m},\frac{\theta_{m}(1-\theta_{m})}{L}\right).$

We then average the estimation over multiple decision periods. When
a user $n$ finishes accessing a channel $m$ for a total $C$ periods,
it updates the estimation of the channel idle probability $\theta_{m}$
as $\tilde{\theta}_{m}(C)=\frac{1}{C}\sum_{i=1}^{C}\hat{\theta}_{m}(i)$,
where $\tilde{\theta}_{m}(C)$ is the estimation of $\theta_{m}$
based on the information of all $C$ decision periods, and $\hat{\theta}_{m}(i)$
is the one-period estimation. By doing so, we have $\tilde{\theta}_{m}(C)\sim\mathcal{N}\left(\theta_{m},\frac{\theta_{m}(1-\theta_{m})}{CL}\right),$
which reduces the variance of one-period MLE by a factor of $C$.

\subsubsection{MLE of Average Data Rate $B_{m}$}

Since the received data rate $b_{n}(t,\tau)$ is also i.i.d over different
time slots and different decision periods, similarly with the MLE
of the channel idle probability $\theta_{m}$, we can obtain the one-period
MLE of mean data rate $B_{m}$ as $\hat{B}_{m}=\frac{\sum_{\tau=1}^{L}b_{n}(t,\tau)}{\sum_{\tau=1}^{L}I_{n}(t,\tau)}$,
and the averaged MLE estimation over $C$ periods as $\tilde{B}_{m}(C)=\frac{1}{C}\sum_{i=1}^{C}\hat{B}_{m}(i).$

By the MLE, we can obtain the estimation of $g(k(t))$, $\theta_{m}$,
and $B_{m}$ as $\tilde{g}(k_{m}(t))$, $\tilde{\theta}_{m}$ and $\tilde{B}_{m}$,
respectively, and then estimate the true expected throughput $U_{n}(t)=\theta_{m}B_{m}g(k(t))$
as $\tilde{U}_{n}(t)=\tilde{\theta}_{m}\tilde{B}_{m}\tilde{g}(k_{m}(t)).$
Since $\tilde{g}(k_{m}(t))$, $\tilde{\theta}_{m}$, and $\tilde{B}_{m}$
follow independent normal distributions with the mean $g(k_{m}(t))$,
$\theta_{m}$, and $B_{m}$, respectively, we thus have $E[\tilde{U}_{n}(t)]=E[\tilde{\theta}_{m}\tilde{B}_{m}\tilde{g}(k_{m}(t))]=U_{n}(t),$
i.e., the estimation of expected throughput $U_{n}(t)$ is unbiased.
In the following analysis, we hence assume that\begin{equation}
\tilde{U}_{n}(t)=U_{n}(t)+\omega_{n},\label{eq:MLE10}\end{equation}
where $\omega_{n}\in(\underline{\omega},\overline{\omega})$ is the
random estimation noise with the probability density function $f(\omega)$
satisfying \begin{eqnarray}
f(\omega)>0,\forall\omega\in(\underline{\omega},\overline{\omega}),\label{eq:MLE11}\\
E[\omega_{n}]=\int_{\underline{\omega}}^{\overline{\omega}}\omega f(\omega)d\omega=0.\label{eq:MLE11-1}\end{eqnarray}

\subsection{Imitative Spectrum Access}

\begin{algorithm}[tt]
\begin{algorithmic}[1]
\State \textbf{initialization:}
\State \hspace{0.4cm} \textbf{choose} a channel $a_{n}$ randomly for each user $n$.
\State \textbf{end initialization\newline}

\Loop{ for each decision period $t$ and each user $n$ in parallel:}
        \For{each time slot $\tau$ in the period $t$}
            \State \textbf{sense} and \textbf{contend} to access the channel $a_{n}$.
            \State \textbf{record} the observations $S_{n}(t,\tau)$, $I_{n}(t,\tau)$ and $b_{n}(t,\tau)$.
        \EndFor
        \State \textbf{estimate} the expected throughput $\tilde{U}_{n}(t)$.
        \State \textbf{select} another user $n'\in\mathcal{N}_{n}$ randomly and \textbf{enquiry} its estimated throughput $\tilde{U}_{n'}(t)$.
        \If{$\tilde{U}_{n'}(t)>\tilde{U}_{n}(t)$}
            \State \textbf{choose} channel $a_{n'}$ (i.e., the one chosen by user $n'$) in the next period.
        \Else{ \textbf{choose} the original channel in the next period.}
        \EndIf

\EndLoop

\end{algorithmic}
\caption{\label{alg:Imitative-Spectrum-Access}Imitative Spectrum Access}
\end{algorithm}

We now propose the imitative spectrum access mechanism in Algorithm \ref{alg:Imitative-Spectrum-Access}. The key motivation is that,  by leveraging the social intelligence of the secondary user crowd, the imitation based mechanism only requires a low computational power for each individual user.
More specifically, we let users imitate the actions of those neighboring users that
achieve a higher throughput (i.e., Lines $11$ to $14$  in Algorithm \ref{alg:Imitative-Spectrum-Access}). This mechanism only relies on local throughput comparisons and is easy to implement in practice.  Each user $n$ at each period $t$ first collects the local observations $\Omega_{n}(t)=\{S_{n}(t,\tau),I_{n}(t,\tau),b_{n}(t,\tau)\}_{\tau=1}^{L}$ (i.e., Lines $5$ to $8$ in Algorithm \ref{alg:Imitative-Spectrum-Access}) and estimates its expected throughput with the MLE method as introduced in Section \ref{MLE} (i.e., Line $9$ in Algorithm \ref{alg:Imitative-Spectrum-Access}). Then user $n$ carries out the imitation by randomly sampling the estimated throughput of another user who shares information with him (i.e., Line $10$ in Algorithm \ref{alg:Imitative-Spectrum-Access}). Such a random sampling can be achieved in different ways. For example, user $n$ can randomly generate a user ID $n'$ from the set $\mathcal{N}_{n}$ and broadcast a throughput enquiry packet including the enquired user ID $n'$. Then user $n'$ will send back an acknowledgement packet including the estimation of its own  expected throughput.

Intuitively, the benefits of adopting the imitation based channel selection are two-fold. On one hand, since each user has incomplete network information, by enquiring another user's throughput information, each user would have a better view of channel environment. If a channel offers a higher data rate, more users trend to exploit the channel due to the nature of imitation. On the other hand, if too many users are utilizing the same channel, a user can improve its data rate through congestion mitigation by imitating users on another channel with less contending users. In the following Section \ref{convergence2}, we show that the proposed imitation-based mechanism can drive a balance between good channel exploitation and congestion mitigation, and achieve a fair spectrum sharing solution.

We shall emphasize that, in the imitative spectrum access mechanism, we require that  each user can  (randomly) select \emph{only one} user for the throughput enquiry, in order to promote diversity in users' channel selections for further congestion mitigation and reduce the system overhead for information exchange. We also evaluate the imitative spectrum access schemes, such that each user can select multiple users for throughput enquiry and imitate the channel selection of the best user among these inquired users (please refer to Section \ref{InquiredUsers} for more details in the separate appendix file). We observe that the performance of  the imitative spectrum access decreases as the number of users for  throughput enquiry increases. This is because when  each user imitates the best  channel selection from multiple users, as the number of enquired users increases, the probability that more users will simultaneously select the same good channel to access in next time slot will increase. This would reduce the diversity of users' channel selections (i.e., increases channel congestion) and hence lead to performance degradation in spectrum sharing, compared with the case of randomly enquiring only one user. Moreover, enquiring multiple users in the same time period will incur a higher system overhead for information exchange.

%% file: Convergence2.tex
\section{\label{convergence2}Convergence of Imitative Spectrum Access}
We then investigate the convergence of imitative spectrum
access. Since users' imitations reply on the information exchange, the structure of the information sharing graph hence plays an important role on the convergence of the mechanism. To better understand the structure property, we will introduce an equivalent and yet more compact  cluster-based graphical representation of the information sharing graph.

\subsection{\label{sub:cluster}Cluster-based Graphical Representation of Information Sharing Graph}
We now introduce the cluster-based presentation. The cluster concept here is similar with the community structure in social networks analysis \cite{scott1988social,alos2008contagion}. Intuitively, a cluster here can be viewed as a set of users
who have similar information sharing structure.  Formally, we define
that

\begin{figure}[tt]
\centering
\includegraphics[scale=0.35]{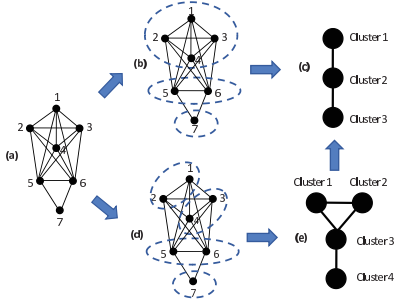}
\caption{\label{fig:An-illustration-of}An illustration of cluster-based representation
of information sharing graph}
\end{figure}

\begin{defn} [\textbf{Cluster}]
A set of users form a cluster if they can share information with each other and they can also
share information with the same set of users that are out of the cluster.
\end{defn}
Taking the information sharing graph on the left hand-side in Figure
\ref{fig:An-illustration-of} as an example, we see that users $1$
to $4$ form a cluster. However, users $1$ to $5$ do not form a
cluster, since user $5$ shares information with user $7$ while user $1$ does not. Furthermore, we can regard a single
user as a special case of cluster. In this case, a general information sharing graph can be represented compactly as a cluster-based graph.
Let $\mathcal{K}=\{1,2,...,K\}$ be the set of clusters, and $w_{kk'}\in\{0,1\}$
denote the information sharing relationship between two clusters $k$
and $k'$. The variable $w_{kk'}=1$ if cluster $k$ communicates
with cluster $k'$ (i.e., the users in cluster $k$ share information
with the users in cluster $k'$) and $w_{kk'}=0$ otherwise. Then we denote the cluster-based graph as $\mathcal{CG}=\{\mathcal{K},\mathcal{W}\}$.
Here vertex set $\mathcal{K}$ is the cluster set, and edge set $\mathcal{W}=\{(k,k'):w_{kk'}=1,\forall k,k'\in\mathcal{K}\}$. We also denote the set of clusters that communicates
with cluster $k$ as $\mathcal{K}_{k}=\{k:(i,k)\in\mathcal{W},\forall k\in\mathcal{K}\}$. Since the users in cluster $k$ and the users in cluster $k'\in \mathcal{K}_{k}$ can  share information with each other, we also define that $\mathcal{C}_{k}\triangleq\mathcal{K}_{k}\cup\{k\}$.

\begin{algorithm}[tt]
\begin{algorithmic}[1]
\State \textbf{$\triangleright$ Construct the set of clusters:}
\State \textbf{set} the un-merged node set $\mathcal{U}=\mathcal{N}$.
\State \textbf{set} cluster index $k=0$.
\Loop{ until $\mathcal{U}=\varnothing$:}
    \State \textbf{select} one node $n\in\mathcal{U}$ randomly.
    \State \textbf{update} cluster index $k=k+1$.
    \State \textbf{set} the set of nodes in cluster $k$ as $\Omega_{k}=\{n\}$.
    \For{each node $m\in\mathcal{N}_{n}\cap\mathcal{U}\backslash\Omega_{k}$}
        \If{$\mathcal{N}_{n}\backslash\{m\}=\mathcal{N}_{m}\backslash\{n\}$}
            \State \textbf{update} $\Omega_{k}=\Omega_{k}\cup\{m\}$.
        \EndIf
    \EndFor
    \State \textbf{update} $\mathcal{U}=\mathcal{U}\backslash\Omega_{k}$.
\EndLoop

\State \textbf{$\triangleright$ Construct the set of edges between clusters:}
\State \textbf{set} the set of $K$ identified clusters above as $\Upsilon=\{1,...,K\}$.
\For{each cluster $k\in\Upsilon$}
    \For{any cluster $h\in\Upsilon\backslash\{k\}$}
        \If{there exists nodes $n\in\Omega_{k}$ and $m\in\Omega_{h}$ such that $m\in\mathcal{N}_{n}$ and $n\in\mathcal{N}_{m}$}
            \State \textbf{set} $w_{kh}=1$.
        \Else{ \textbf{set} $w_{kh}=0$.}
        \EndIf
    \EndFor
\EndFor
\end{algorithmic}
\caption{\label{alg:Cluster}Algorithm for Constructing Cluster-based Graph}
\end{algorithm}

As illustrated in Figure \ref{fig:An-illustration-of}, an information sharing graph can be represented as different cluster-based graphs (e.g., graphs (c) and (e) in Figure \ref{fig:An-illustration-of}). In general fact, we can first consider the original information sharing graph as a primitive cluster-based graph by regarding each single user as a cluster (e.g., graph (a) in Figure \ref{fig:An-illustration-of}). We then further carry out the clustering (e.g., graph (d) in Figure \ref{fig:An-illustration-of}) and obtain the cluster-based graph (e). We next merge clusters $1$ and $2$ of this cluster-based graph into one cluster and then obtain the most compact cluster-based graph (c) in this example. We summarize the algorithm for constructing the cluster-based graph in Algorithm \ref{alg:Cluster}. Note that for the practical implementation, the knowledge of cluster-based graphs is not required. The use of clutter-based graph here is to facilitate the analysis of the convergence of imitative spectrum access mechanism. The convergence properties of the imitative spectrum access mechanism are determined by the original information exchange graph, and are the same for all these cluster-based graphs since they preserve the structural property of the original information exchange graph (i.e., two users share information in the cluster-based graph if and only if they share information in the original graph).

Next we explore the property of the cluster-based representation. We denote the cluster that a user $n\in\mathcal{N}$
belongs to as $k(n)$ and the set of users in cluster $k\in\mathcal{K}$
as $\mathcal{N}(k)$. According to the definition of cluster, we can see that if users $n$ and $n'$ share information with each other, then they either belong to the same cluster or two different clusters that communicate with each other. Thus we have that

\begin{lem}
\label{lem:The-set-of}The set of users that share information with
user $n$ is the same as the set of users in user $n$'s cluster and the clusters that communicate
with user $n$'s cluster, i.e., $\mathcal{N}_{n}=\cup_{k'\in\mathcal{C}_{k(n)}}\mathcal{N}(k')$.
\end{lem}
Furthermore, the cluster-based representation also preserves
the connectivity of the information sharing graph (i.e., it is possible to find a path from any node to any other node).
\begin{lem}
\label{lem:If-the-information}The information sharing graph is
connected if and only if the corresponding cluster-based graph is also connected.
\end{lem}

\begin{figure}[tt]
\centering
\includegraphics[scale=0.6]{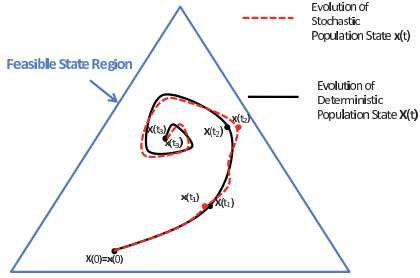}
\caption{\label{fig:Illustration-of-the}Illustration of the approximation
of stochastic population state $\boldsymbol{x}(t)$ by deterministic
population state $\boldsymbol{X}(t)$}
\end{figure}

\subsection{Dynamics of Imitative Spectrum Access}

Based on the cluster-based graphical representation of information sharing graph, we next study the evolution dynamics
of the imitative spectrum access mechanism. Suppose that the underlying information sharing graph can be represented
by $K$ clusters, and the number of users in each cluster $k$ is $z_{k}$ with $\sum_{k=1}^{K}z_{k}=N$. For the ease of exposition,
we will focus the case that the number of users $z_{k}$ in each cluster $k$ is large. Numerical
results show that the observations also hold for the case that the
number of  users in a cluster is small (see Sections \ref{sec:Simulation-Results} for details).

With a large cluster user population,
it is convenient to use the population state $\boldsymbol{x}(t)$
to describe the dynamics of spectrum access. We then denote the population
state of all users as $\boldsymbol{x}(t)\triangleq(\boldsymbol{x}^{1}(t),...\boldsymbol{x}^{K}(t))$
and the population state of cluster $k$ as $\boldsymbol{x}^{k}(t)\triangleq(x_{1}^{k}(t),...,x_{M}^{k}(t))$. Here $x_{m}^{k}(t)$ denotes the fraction of users in cluster $k$
choosing channel $m$ to access at period $t$, and we have $\sum_{m=1}^{M}x_{m}^{k}(t)=1$.

In the imitative spectrum access mechanism, each user $n$ relies
on its local estimated expected throughput $\tilde{U}_{n}(t)$ to
decide whether to imitate other user's channel selection. Due to the
random estimation noise $\omega_{n}$, the evolution of the population
state $\{\boldsymbol{x}(t),\forall t\geq0\}$ is stochastic and difficult to
analyze directly. However, when the population of cluster users $z_{k}$ is large,
due to the law of large number, such stochastic process can be well
approximated by its mean deterministic trajectory $\{\boldsymbol{X}(t),\forall t\geq0\}$
\cite{key-IM3}. Here $\boldsymbol{X}(t)\triangleq(\boldsymbol{X}^{1}(t),...\boldsymbol{X}^{K}(t))$
is the deterministic population state of all the users, and $\boldsymbol{X}^{k}(t)\triangleq(X_{1}^{k}(t),...,X_{M}^{k}(t))$
is the deterministic population state of cluster $k$. Consider a user in cluster $k$ chooses channel $i$ , and let $P_{i,k}^{j}(\boldsymbol{X}(t))$
denote the probability that this user in the deterministic population state $\boldsymbol{X}(t)$ will
choose channel $j$ in next period. According to \cite{key-IM3}, we
have
\begin{lem}
\label{lem:There-exists-a}There exists a scalar $\delta$ such
that, for any bound $\epsilon>0$, decision period $T>0$, and any
large enough cluster size $z_k$, the maximum difference between
the stochastic and deterministic population states over all periods
is upper-bounded by $\epsilon$ with an arbitrarily large probability,
i.e., \begin{equation}
Pr\{\max_{0\leq t\leq T}\max_{m\in\mathcal{M}}|X_{m}^{k}(t)-x_{m}^{k}(t)|\geq\epsilon\}\leq e^{-\epsilon^{2}z_k}, \forall k\in\mathcal{K}, \label{eq:DD1}\end{equation}
given that $\boldsymbol{X}(0)=\boldsymbol{x}(0)$.
\end{lem}

The proof is similar with Lemma $1$ in \cite{key-IM3} and hence is omitted
here. As illustrated in Figure \ref{fig:Illustration-of-the}, Lemma
\ref{lem:There-exists-a} indicates that the trajectory of the stochastic
population state $\{\boldsymbol{x}(t),\forall t\geq0\}$ is within
a small neighborhood of the trajectory of the deterministic population
state $\{\boldsymbol{X}(t),\forall t\geq0\}$ when the user population
$N$ is large enough. Moreover, since the MLE is unbiased, the deterministic
population state $\{\boldsymbol{X}(t),\forall t\geq0\}$ is also the
mean field dynamics of stochastic population state $\{\boldsymbol{x}(t),\forall t\geq0\}$
\cite{key-IM3}. If the deterministic dynamics $\{\boldsymbol{X}(t),\forall t\geq0\}$
converge to an equilibrium, the stochastic dynamics $\{\boldsymbol{x}(t),\forall t\geq0\}$
must also converge to the same equilibrium on the time average \cite{key-IM3}.

We now study the evolution dynamics of the deterministic population
state $\{\boldsymbol{X}(t),\forall t\geq0\}$.
Let $U(m,\boldsymbol{X}(t))=\theta_{m}B_{m}g\left(\sum_{k=1}^{K}z_{k}X_{m}^{k}(t)\right)$
denote the expected throughput of a user that chooses channel $m$
with a total of $\sum_{k=1}^{K}z_{k}X_{m}^{k}(t)$ contending users
in the population state $\boldsymbol{X}(t)$. Recall that in the imitative
spectrum access mechanism, each user will randomly choose another
user that shares information with it, and imitate that user's channel selection
if that user's estimated throughput is higher. Suppose that the user $n$ is in cluster
$k$ choosing channel $i$. According to Lemma \ref{lem:The-set-of}, the set of users that share information with user $n$ are in set of clusters $\mathcal{C}_{k}$.
Thus, we can obtain the probability $P_{i,k}^{j}(\boldsymbol{X}(t))$
that this user $n$ will imitate
another user $n'$ on channel $j$ in next period as\begin{align}
P_{i,k}^{j}(\boldsymbol{X}(t))= & \sum_{k'\in\mathcal{C}_{k}}\frac{z_{k'}}{\sum_{l\in\mathcal{C}_{k}}z_{l}}X_{j}^{k'}(t)\nonumber\\
& \times Pr\{\tilde{U}(j,\boldsymbol{X}(t))>\tilde{U}(i,\boldsymbol{X}(t))\}.\label{eq:DD3}\end{align}
Here $\frac{z_{k'}}{\sum_{l\in\mathcal{C}_{k}}z_{l}}X_{j}^{k'}(t)$
denotes the probability that a user choosing channel $j$ in cluster
$k'\in\mathcal{C}_{k}$ will be selected for imitation. From (\ref{eq:MLE10}), we have\begin{align}
\tilde{U}(j,\boldsymbol{X}(t))-\tilde{U}(i,\boldsymbol{X}(t))  & = U(j,\boldsymbol{X}(t))-U(i,\boldsymbol{X}(t)) \nonumber \\
& +\omega_{n'}-\omega_{n},\label{eq:DD4}\end{align}
where $\omega_{n},\omega_{n'}$ are the random estimation noises with
the probability density function $f(\omega)$. Let $\varpi=\omega_{n}-\omega_{n'}$,
and we can obtain the probability density function of random variable
$\varpi$ as\begin{equation}
q(\varpi)=\int_{\underline{\omega}}^{\overline{\omega}}f(\omega)f(\varpi+\omega)d\omega.\label{eq:DD5}\end{equation}
We further denote the cumulative distribution function $\varpi$ as
$Q(\varpi)$, i.e., $Q(\varpi)=\int_{-\infty}^{\varpi}q(s)ds$. Then
from (\ref{eq:DD3}) and (\ref{eq:DD4}), we have for any $j\neq i$,\begin{equation}
P_{i,k}^{j}(\boldsymbol{X}(t))=\sum_{k'\in\mathcal{C}_{k}}\frac{z_{k'}}{\sum_{l\in\mathcal{C}_{k}}z_{l}}X_{j}^{k'}(t)Q(U(j,\boldsymbol{X}(t))-U(i,\boldsymbol{X}(t))),\label{eq:DD6}\end{equation}
and \begin{equation}
P_{i,k}^{i}(\boldsymbol{X}(T))=1-\sum_{j\neq i}P_{i,k}^{j}(\boldsymbol{X}(t)).\label{eq:DD7}\end{equation}
Based on (\ref{eq:DD6}) and (\ref{eq:DD7}), we
obtain the evolution dynamics of the deterministic population state
$\{\boldsymbol{X}(t),\forall t\geq0\}$ as (the proof is given in Section \ref{proof2} in the separate appendix file)
\begin{thm} \label{thm1}
For the imitative spectrum access mechanism, the evolution dynamics of deterministic population
state $\{\boldsymbol{X}(t),\forall t\geq0\}$ are given as\begin{align}
\dot{X}_{m}^{k}(t) & =  \sum_{i=1}^{M}X_{i}^{k}(t)\sum_{k'\in\mathcal{C}_{k}}\frac{z_{k'}}{\sum_{l\in\mathcal{C}_{k}}z_{l}}X_{m}^{k'}(t) \nonumber \\
& \times Q(U(m,\boldsymbol{X}(t))-U(i,\boldsymbol{X}(t)))\nonumber \\
 &  -X_{m}^{k}(t)\sum_{i=1}^{M}\sum_{k'\in\mathcal{C}_{k}}\frac{z_{k'}}{\sum_{l\in\mathcal{C}_{k}}z_{l}}X_{i}^{k'}(t)\nonumber \\
 & \times Q(U(i,\boldsymbol{X}(t))-U(m,\boldsymbol{X}(t))),\label{eq:DD8}\end{align}
where the derivative is with respect to time $t$.\end{thm}

\subsection{Convergence of Imitative Spectrum Access}
We now study the convergence of the imitative spectrum access mechanism.
Let $\boldsymbol{x}^{*}\triangleq(\boldsymbol{x}^{1*},...,\boldsymbol{x}^{K*})$
denote the equilibrium of the imitative spectrum access, and $a_{n}^{*}$
denote the channel chosen by user $n$ in the equilibrium $\boldsymbol{x}^{*}$.
We first introduce the definition of \emph{imitation equilibrium. }
\begin{defn} [\textbf{Imitation Equilibrium}]
\label{def:A-population-state-1}A population state $\boldsymbol{x}^{*}$
is an imitation equilibrium if and only if for each user $n\in\mathcal{N}$
,\begin{equation}
U(a_{n}^{*},\boldsymbol{x}^{*})\geq\max_{a\in\Delta_{n}(\boldsymbol{x}^{*})\backslash\{a_{n}^{*}\}}U(a,\boldsymbol{x}^{*}),\label{eq:IM2-1}\end{equation}
where $\Delta_{n}(\boldsymbol{x}^{*})\triangleq\{m\in\mathcal{M}:\exists a_{i}^{*}=m,\forall i\in\mathcal{N}_{n}\}$
denotes the set of channels are chosen by users that share information
with user $n$ in the equilibrium $\boldsymbol{x}^{*}$.
\end{defn}
The intuition of Definition \ref{def:A-population-state-1} is that
no imitation can be carried out to improve any user's data rate in the equilibrium.  For the imitative spectrum access mechanism, we show that
\begin{thm}
\label{thm:For-the-imitative-1}For the imitative spectrum access
mechanism, the evolution dynamics of deterministic population state
$\{\boldsymbol{X}(t),\forall t\geq0\}$ asymptotically converge to
an imitation equilibrium $\boldsymbol{X}^{*}$such that\begin{align}
U(m,\boldsymbol{X}^{*})=U(i,\boldsymbol{X}^{*}),\forall m,i\in\Delta_{n}(\boldsymbol{X}^{*}),\forall n\in\mathcal{N}.\label{eq:IM2}\end{align}
\end{thm}

The proof is given in Section \ref{proof1} in the separate appendix file. According to Lemma \ref{lem:There-exists-a}, we know that the
stochastic imitative spectrum access dynamics $\{\boldsymbol{x}(t),\forall t\geq0\}$
will be attracted into a small neighborhood around the imitation
equilibrium $\boldsymbol{X}^{*}$. Moreover, since the imitation equilibrium
$\boldsymbol{X}^{*}$ is also the mean field equilibrium of stochastic
dynamics $\{\boldsymbol{x}(t),\forall t\geq0\}$, the stochastic dynamics $\{\boldsymbol{x}(t),\forall t\geq0\}$
hence converge to the imitation equilibrium $\boldsymbol{X}^{*}$
on the time average. That is, the fraction of users adopting a certain channel selection will converge to a fixed vale on the time average. However, a user would keep switching its channel during the process. This is because that when many other users also utilize the same channel, the user would imitate to select another channel with less contending users to mitigate congestion. The mechanism hence can drive a balance between good channel exploitation and congestion mitigation.

According to the definition of $\Delta_{n}(\boldsymbol{X}^{*})$, we see from Theorem \ref{thm:For-the-imitative-1} that two users will achieve the same expected throughput if they share information with each other (i.e., they are neighbors in the information sharing graph). Moreover, when the information sharing graph is connected, we can show that all the users achieve the same throughput at the imitation equilibrium.
\begin{cor}\label{cor222}
When the information sharing graph is connected, all the users
following the imitative spectrum access mechanism achieve the same
expected throughput, i.e., $U(a_{n}^{*},\boldsymbol{X}^{*})=U(a_{n'}^{*},\boldsymbol{X}^{*}),\forall n,n'\in\mathcal{N}.$\end{cor}
%\begin{proof}
%Since the information sharing graph is connected, for any two different
%users $n$ and $n'$, there must exists a path $(n_{1}=n,n_{2},...,n_{L}=n')$
%satisfying that $n_{l+1}\in\mathcal{N}_{n_{l}},\forall1\leq l\leq L-1$.
%According to Theorem \ref{thm:For-the-imitative-1}, we have $U(m,\boldsymbol{X}^{*})=U(i,\boldsymbol{X}^{*}),\forall m,i\in\Delta_{n}(\boldsymbol{X}^{*}),\forall n\in\mathcal{N}.$ Since $a_{n_{l+1}}^{*}\in\Delta_{n_{l}}(\boldsymbol{X}^{*}),\forall1\leq l\leq L-1$, it implies that $U(a_{n_{1}}^{*},\boldsymbol{X}^{*})=U(a_{n_{2}}^{*},\boldsymbol{X}^{*})=...=U(a_{n_{L}}^{*},\boldsymbol{X}^{*}).$ \end{proof}
The proof is given in Section \ref{proof22} in the separate appendix file. Furthermore, we can show in Corollary \ref{cor333}  that the convergent imitation equilibrium is the most fair channel allocation in terms of the widely-used Jain's fairness index $J=\frac{(\sum_{n=1}^{N} U(a_{n}^{*},\boldsymbol{X}^{*}))^{2}}{N\sum_{n=1}^{N}U(a_{n}^{*},\boldsymbol{X}^{*})^{2}}$) \cite{jain1984quantitative}. Notice that the fair channel allocation is due to the nature of imitation. If the channel allocation is unfair, there must exist some secondary users that achieve a higher throughput  than others. In this case, other users with a lower throughput will imitate the channel selection of those users until the performance of all users are equal (i.e., fair spectrum sharing).
\begin{cor}\label{cor333}
When the information sharing graph is connected, the Jain's fairness index $J$ is maximized at the imitation equilibrium.\end{cor}
%\begin{proof}
%According to Cauchy-Schwarz inequality, we know that
%$\left(\sum_{n=1}^{N}U(a_{n}^{*},\boldsymbol{X}^{*})\right)^{2}\leq N\sum_{n=1}^{N}U(a_{n}^{*},\boldsymbol{X}^{*})^{2}$
%and $\left(\sum_{n=1}^{N}U(a_{n}^{*},\boldsymbol{X}^{*})\right)^{2}=N\sum_{n=1}^{N}U(a_{n}^{*},\boldsymbol{X}^{*})^{2}$
%if and only if $U(a_{n}^{*},\boldsymbol{X}^{*})=U(a_{m}^{*},\boldsymbol{X}^{*})$,for any $n,m=1,...,N$.
%It then follows that Jain's fairness index $J$ is maximized at the imitation
%equilibrium, since the condition $U(a_{n}^{*},\boldsymbol{X}^{*})=U(a_{m}^{*},\boldsymbol{X}^{*})$
%holds according to Corollary \ref{cor222}.
%\end{proof}
The proof is given in Section \ref{proof33} in the separate appendix file. We next discuss the efficiency of the imitation equilibrium. Similar to the definition of price of anarchy
(PoA) in game theory, we will quantify the efficiency ratio of imitation
equilibrium $\boldsymbol{X}^{*}$ over the centralized optimal solution
and define the price of imitation (PoI) as
\[
\mbox{PoI}=\frac{\sum_{n=1}^{N}U(a_{n}^{*},\boldsymbol{X}^{*})}{\max_{\boldsymbol{X}}\sum_{n=1}^{N}U(a_{n},\boldsymbol{X})}.
\]
Since it is difficult to analytically characterize the PoI for the general case, we focus
on the case that all the channels are homogenous, i.e., $B_{m}=B_{m'}=B$
and $\theta_{m}=\theta_{m'}=\theta$ for any $m,m'=1,...,M$. Let
$Z$ be the number of channels being utilized at the imitation equilibrium
$\boldsymbol{X}^{*}$, i.e., $Z=|\cup_{n=1}^{N}\Delta_{n}(\boldsymbol{X}^{*})|$.
We can show the following result.
\begin{thm}\label{PoI}
When the information sharing graph is connected and all the channels
are homogenous, the PoI of imitative spectrum access mechanism is
at least $\frac{Ng(\frac{N}{Z})}{M}.$\end{thm}
%\begin{proof}
%First of all, according to Corollary \ref{cor222}, we know that all the users
%at the imitation equilibrium $\boldsymbol{X}^{*}$achieve the same
%throughput. Since all the channels are homogenous, the number of users
%on each of $Z$ utilized channels is the same, $\frac{N}{Z}$. It
%then follows that the system-wide throughput at the imitation equilibrium
%is $\sum_{n=1}^{N}U(a_{n}^{*},\boldsymbol{X}^{*})=NB\theta g(\frac{N}{Z}).$
%On the other hand, for the centralized optimal solution, since $kg(k)\leq1$
%for any $k=1,2,...,N$, we know that $\max_{\boldsymbol{X}}\sum_{n=1}^{N}U(a_{n},\boldsymbol{X})=\max_{(k_{1},..,k_{M})}\sum_{m=1}^{M}k_{m}B\theta g(k_{m})\leq MB\theta$.
%Thus, we can conclude that the PoI is at least $\frac{Ng(\frac{N}{Z})}{M}$.
%\end{proof}
The proof is given in Section \ref{proof44} in the separate appendix file. We also evaluate the performance of imitative spectrum access mechanism
for the general case in Section \ref{sec:Simulation-Results}. Numerical results demonstrate that
the mechanism is efficient, with at most $20\%$ performance loss,
compared with the centralized optimal solution.

%% file: Innovation.tex
\section{Imitative Spectrum Access With User Heterogeneity}\label{ISA2}

For the ease of exposition, we have considered the case that users
are homogeneous, i.e., different users achieve the same data
rate on the same channel. We now consider the general heterogeneous case where different users may achieve different data rates on the same
channel.

Let $b_{m}^{n}(\tau)$ be the realized data rate of user $n$ on an
idle channel $m$ at a time slot $\tau$, and $B_{m}^{n}$ be the
mean data rate of user $n$ on the idle channel $m$, i.e., $B_{m}^{n}=E[b_{m}^{n}(\tau)]$.
In this case, the expected throughput of user $n$ is given as $U_{n}^{m}=\theta_{m}B_{m}^{n}g(k_{m})$.
For imitative spectrum access mechanism in Algorithm \ref{alg:Imitative-Spectrum-Access}, each user
carries out the channel imitation by comparing its throughput with
the throughput of another user. However, such throughput comparison
may not be feasible when users are heterogeneous, since a user may
achieve a low throughput on a channel that offers a high throughput for
another user.

To address this issue, we propose a new imitative spectrum access
mechanism with user heterogeneity in Algorithm \ref{alg:Imitative-Spectrum-Access2}. More specifically, when
a user $n$ on a channel $m$ randomly selects another neighboring user $n'$
on another channel $m'$, user $n'$ informs user $n$ about the estimated
channel grabbing probability $\tilde{g}(k_{m'})$ instead of the estimated expected throughput. Then user $n$
will compute the estimated expected throughput on channel
$m'$ as \begin{equation}
\tilde{U}_{n}^{m'}=\tilde{\theta}_{m'}\tilde{B}_{m^{'}}^{n}\tilde{g}(k_{m^{'}}).\label{eq:im2222}\end{equation}
 If $\tilde{U}_{n}^{m'}>\tilde{U}_{n}^{m}$, then user $n$ will imitate the channel
selection of user $n'$.

To implement the mechanism above, each user $n$ must have the information
of its own estimated channel idle probability $\tilde{\theta}_{m'}$ and
data rate $\tilde{B}_{m^{'}}^{n}$ of the unchosen channel $m'$.
Hence we add an initial channel estimation stage in the imitative
spectrum access mechanism in Algorithm \ref{alg:Imitative-Spectrum-Access2}. In this stage, each user
initially estimates the channel idle probability $\tilde{\theta}_{m}$
and data rate $\tilde{B}_{m}^{n}$ by accessing all the channels in
a randomized round-robin manner. This ensures that all users do not
choose the same channel at the same period. Let $\mathcal{M}_{n}$
(equals to the empty set $\oslash$ initially) be set of channels probed by user
$n$ and $\mathcal{M}_{n}^{c}=\mathcal{M}\backslash\mathcal{M}_{n}$.
At beginning of each decision period, user $n$ randomly chooses a
channel $m\in\mathcal{M}_{n}^{c}$ (i.e., a channel that has not been
accessed before) to access. At end of the period, user $n$ can estimate
the channel idle probability $\tilde{\theta}_{m}$ and data rate $\tilde{B}_{m}^{n}$
according to the MLE method introduced in Section \ref{MLE}.

Numerical results show that the proposed imitative spectrum access
mechanism with user heterogeneity can still converge to an imitation equilibrium satisfying the definition in (\ref{def:A-population-state-1}), i.e., no user can further improve
its expected throughput by imitating another user. Numerical results show that the imitative spectrum access mechanism with user heterogeneity achieves up-to $500\%$  fairness improvement with at most $20\%$ performance loss, compared with the centralized optimal solution. This demonstrates that the proposed imitation-based mechanism can achieve efficient spectrum utilization and meanwhile provide good fairness across secondary users.

%Note that when the set of chosen channels includes all channels, i.e., $\Delta_{n}(\boldsymbol{X}^{*})=\mathcal{M},$
%it is easy to verify that the imitation equilibrium $\boldsymbol{X}^{*}$
%is a Nash equilibrium, since no user can improve its payoff by changing
%its channel unilaterally. Numerical results show that the imitative spectrum access mechanism with user heterogeneity
%converges to a Nash equilibrium by utilizing all the channels. Moveover, the mechanism achieves up-to $530\%$  fairness improvement with at most $20\%$ performance loss, compared with the centralized optimal solution.

\begin{algorithm}[tt]
\begin{algorithmic}[1]
\Loop{ for each user $n\in\mathcal{N}$ in parallel:\newline\Comment{\textbf{\emph{Initial Channel Estimation Stage}}}}

    \While{$\mathcal{M}_{n}\neq\mathcal{M}$}
        \State \textbf{choose} a channel $m$ from the set $\mathcal{M}_{n}^{c}$ randomly.
        \State \textbf{sense} and \textbf{contend} to access the channel $m$ at each time slot of the decision period.
        \State \textbf{record} the observations $S_{n}(t,\tau)$, $I_{n}(t,\tau)$ and $b_{n}(t,\tau)$.
        \State \textbf{estimate} the channel idle probability $\tilde{\theta}_{m}$ and data rate $\tilde{B}_{m}^{n}$.
        \State \textbf{set} $\mathcal{M}_{n}=\mathcal{M}_{n}\cup\{m\}$.
    \EndWhile{\newline\newline\Comment{\textbf{\emph{Imitative Spectrum Access Stage}}}}

    \For{each time period $t$}
        \State \textbf{sense} and \textbf{contend} to access the channel $m$ at each time slot of the decision period.
        \State \textbf{record} the observations $S_{n}(t,\tau)$, $I_{n}(t,\tau)$ and $b_{n}(t,\tau)$.
        \State \textbf{estimate} the expected throughput $\tilde{U}_{n}^{a_{n}}(t)$.
        \State \textbf{select} another user $n'\in\mathcal{N}_{n}$ randomly and \textbf{enquiry} its channel grabbing probability $\tilde{g}(k_{a_{n'}})$.
        \State \textbf{estimate} the expected throughput $\tilde{U}_{n}^{a_{n'}}(t)$ based on (\ref{eq:im2222}).
        \If{$\tilde{U}_{n}^{a_{n'}}(t)>\tilde{U}_{n}^{a_{n}}(t)$}
            \State \textbf{choose} channel $a_{n'}$ (i.e., the one chosen by user $n'$) in the next period.
        \Else{ \textbf{choose} the original channel in the next period.}
        \EndIf
    \EndFor
\EndLoop

\end{algorithmic}
\caption{\label{alg:Imitative-Spectrum-Access2}Imitative Spectrum Access With User Heterogeneity}
\end{algorithm}

%% file: Results.tex
%!TEX root = main.tex
%SourceDoc main.tex

\section{\label{sec:Simulation-Results}Simulation Results}

In this section, we evaluate the proposed imitative spectrum access
mechanisms by simulations. We consider a spectrum sharing network consisting
$M=5$ Rayleigh fading channels.
The data rate on an idle channel $m$ of user $n$ is computed according to the
Shannon capacity, i.e., $b_{m}^{n}=E_{m}\log_{2}(1+\frac{\eta_{n}h_{m}^{n}}{n_{0}}),$
where $E_{m}$ is the bandwidth of channel $m$, $\eta_{n}$ is
the power adopted by user $n$, $n_{0}$ is the noise power, and $h_{m}^{n}$
is the channel gain (a realization of a random variable that follows
the exponential distribution with the mean $\bar{h}_{m}^{n}$). By setting different mean channel gain $\bar{h}_{m}^{n}$, we can have different mean data rates $B_{m}^{n}=E[b_{m}^{n}]$. In the
following simulations, we set $\zeta_{m}=10$ MHz, $n_{0}=-100$ dBm,
and $\eta_{n}=100$ mW. We set the number of time slots in each decision
period as $100$. We will consider both cases with homogeneous and heterogenous users.

%For the expected throughput estimation, we implement the maximum likelihood estimation
%(MLE) with different period length
%$L$ in Figure \ref{fig:MLE}. For each fixed $L$, we repeat the MLE for $50$ times and
%calculate the mean and variance of the estimated values. We see that as
%the number of observations increases, the estimated mean converges to the true value and the variance of the estimated value decreases. In following simulations, we set the period length $L=200$, which can achieve a good estimation of the expected throughput.
%
%
%\begin{figure}[tt]
%\centering
%\includegraphics[scale=0.5]{MLE}
%\caption{\label{fig:MLE}Maximum likelihood estimation of the expected throughput. Vertical bar represents
%the range of the estimated values.}
%\end{figure}

\subsection{\label{SG}Imitative Spectrum Access with Homogeneous Users}

\subsubsection{\label{SG1}I.i.d. Channel Environment}

\begin{figure}[tt]
\centering
\includegraphics[scale=0.3]{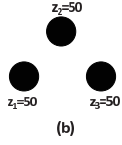}
\caption{\label{fig:Four-different-types}four types of cluster-based
graphs with $z_{k}$ representing the number of users in each cluster
$k$}
\end{figure}

\begin{figure}[tt]
\centering
\includegraphics[scale=0.45]{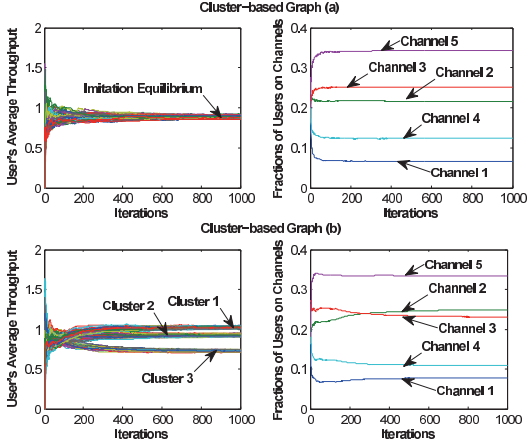}
\caption{\label{fig:Users'-average-throughputs}Users' average throughputs and fractions of users on different channels
on cluster-based graphs (a) and (b) in Figure \ref{fig:Four-different-types}}
\end{figure}

%\begin{figure}[tt]
%\centering
%\includegraphics[scale=0.5]{fig_cd}
%\caption{\label{fig:Users'-average-throughputs2}Users' average throughputs and fractions of users on different channels
%on cluster-based graphs (c) and (d) in Figure \ref{fig:Four-different-types}}
%\end{figure}

We first implement the imitative spectrum access mechanism  with  $N=150$ homogeneous users (i.e., Algorithm \ref{alg:Imitative-Spectrum-Access}) and the number of backoff mini-slots $\lambda_{\max}=50$. For each user $n$, the mean channel data rates $\{B_{m}^{n}\}_{m=1}^{M}=\{15,70,90,40,100\}$
Mbps, respectively. The channel states are i.i.d. Bernoulli random variable with the mean idle probabilities $\{\theta_{m}\}_{m=1}^{M}=\{\frac{2}{3},\frac{4}{7},\mbox{\ensuremath{\frac{5}{9}},\ensuremath{\frac{1}{2},\frac{4}{5}}}\}$, respectively.

We consider that the information sharing graphs are represented by
different cluster-based graphs as shown in Figure \ref{fig:Four-different-types}.
In Graph (a), clusters $1$ and $3$ do not communicate directly and
they are connected to cluster $2$. In Graph (b), all three clusters
are isolated. We show the time average user's throughput
in Figure \ref{fig:Users'-average-throughputs}. We see that all
the users achieve the same average throughput on Graph (a).
This verifies the theoretic result that when information sharing graph is connected (i.e., the corresponding cluster-based graph is connected), all the users achieve the same average
throughput in the imitation equilibrium. When information sharing graph is not connected (e.g., Graph (b)), we see that users in different
clusters may achieve different throughputs. However, all the users
in the same cluster have the same average throughput. This is also
an imitation equilibrium given the constraint of their information
sharing. Moreover, we see that all the channels will be utilized in the imitation equilibria on both Graphs (a)and (b). A channel of a higher data rate will be utilized by a larger fraction of users.

\subsubsection{\label{MCE}Markovian Channel Environment}
\begin{figure}[tt]
\centering
\includegraphics[scale=0.35]{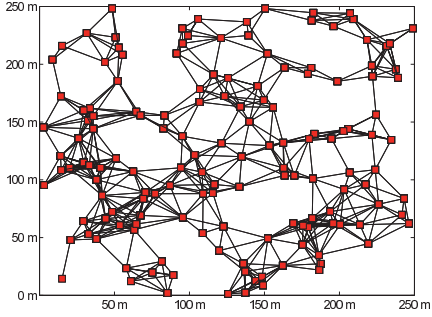}
\caption{\label{fig:A-square-area}A square area of a length of $250$ m with
$150$ scattered users. Each user can share information with those users that
are connected with it by an edge.}
\end{figure}

For the interests of obtaining closed form solutions and deriving engineering insights, we have considered the i.i.d. channel
model so far. We now evaluate the proposed mechanism in the Markovian
channel environment. We denote the channel state probability
vector of channel $m$ at a time slot $\tau$ as $\boldsymbol{p}_{m}(\tau)\triangleq(Pr\{S_{m}(\tau)=0\},Pr\{S_{m}(\tau)=1\}),$
which follows a two-states Markov chain as $\boldsymbol{p}_{m}(\tau+1)=\boldsymbol{p}_{m}(\tau)\Gamma_{m},\forall \tau\geq1$,
with the transition matrix $\Gamma_{m}=\left[\begin{array}{cc}
1-p_{m} & p_{m}\\
q_{m} & 1-q_{m}\end{array}\right].$ In this case, we can obtain the stationary distribution that the
channel $m$ is idle with a probability of $\theta_{m}=\frac{p_{m}}{p_{m}+q_{m}}$. The study in \cite{lopez2011empirical} shows that the statistical properties of spectrum usage from empirical measurement data can be accurately captured and reproduced by properly setting the transition matrix. In this experiment, we choose different $p_{m}$ and $q_{m}$ for different channels such that the idle probabilities $\{\theta_{m}\}_{m=1}^{M}=\{\frac{2}{3},\frac{4}{7},\mbox{\ensuremath{\frac{5}{9}},\ensuremath{\frac{1}{2},\frac{4}{5}}}\}$
are the same as before. We consider that $N=150$ users are randomly scattered
across a square area of a side-length of $250$ m with the information sharing graph as shown in Figure \ref{fig:A-square-area}. As mentioned in Section \ref{sub:cluster}, this information sharing graph  can also be regarded as a cluster-based graph by considering a single user as a cluster.

\begin{figure}[tt]
\centering
\includegraphics[scale=0.45]{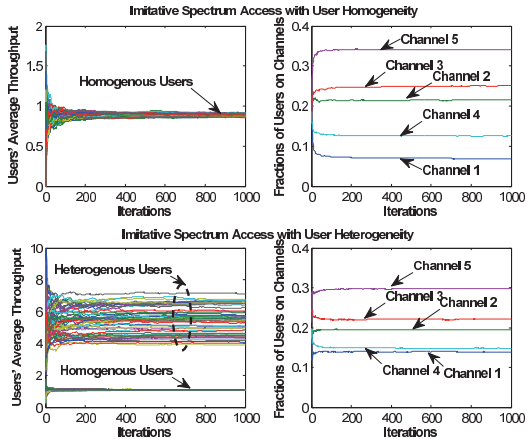}
\caption{\label{fig:Users'-average-throughputs-1}Users' average throughputs and fractions of users on different channels
on the information sharing graph in Figure \ref{fig:A-square-area}}
\end{figure}

The results are shown in the upper part of Figure \ref{fig:Users'-average-throughputs-1}. We observe that the imitative spectrum access mechanism still achieves the imitation equilibrium
in the Markovian channel environment. The average throughput that each user achieves is the same as that in i.i.d. channel environment on the connected graph (a) in Figures \ref{fig:Users'-average-throughputs}. This is because that our proposed
Maximum Likelihood Estimation of the channel idle probability $\theta_{m}$
follows the sample average approach. By the law of large
numbers, when the observation samples are sufficient, such a sample
average approach can achieve an accurate estimation of the average
statistics of the channel availability, even if the channel state
is not an i.i.d. process.

\begin{figure}[tt]
\centering
\includegraphics[scale=0.45]{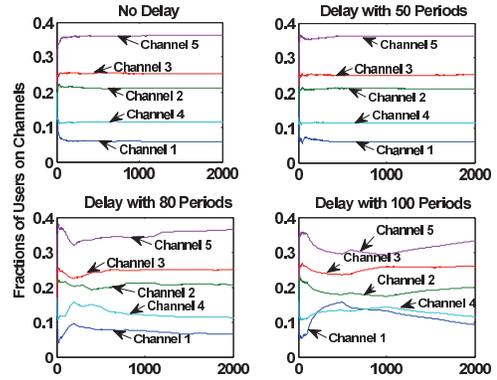}
\caption{\label{fig:Delay}Imitative spectrum access with information exchange delay.}
\end{figure}

\subsection{\label{SG}Imitative Spectrum Access with Heterogeneous Users}

We then implement the imitation spectrum access mechanism with heterogeneous users (i.e., Algorithm \ref{alg:Imitative-Spectrum-Access2}) on the same information sharing graph in Figure \ref{fig:A-square-area}. The mean data rates of $100$ randomly chosen users out of these $150$ users  are homogeneous with the same data rates as before (i.e., $\{B_{m}^{n}\}_{m=1}^{M}=\{15,70,90,40,100\}$ Mbps). For the remaining $50$ users,  we set that the users' data rates are heterogenous with the mean data rate of user $n$ on channel $m$ as $B_{m}^{n}=100+R$ where $R$ is a random value drawn from the uniform distribution over $(0,100)$. The results are shown in the bottom part of Figure \ref{fig:Users'-average-throughputs-1}. We see that the mechanism converges to the equilibrium wherein homogeneous users achieve the same expected throughput and heterogenous users may achieve different expected throughputs. Moreover, we observe that the mechanism converges to a stable user distribution on channels. This implies that no user can further improve its expected throughput by imitating another user. That is, the equilibrium is an imitation equilibrium satisfying the definition in (\ref{eq:IM2-1}).

\subsection{\label{Delay}The Impact of Information Exchange Delay}
Next we  evaluate the impact of the information exchange delay on the imitative spectrum access mechanism. Similar as  in Section \ref{MCE}, we consider that the spectrum sharing network consists of $M=5$ Rayleigh fading channels with the mean data rates $\{B_{m}^{n}\}_{m=1}^{M}=\{15,70,90,40,100\}$ Mbps, respectively. The primary activities are Markovian with the channel idle probabilities $\{\theta_{m}\}_{m=1}^{M}=\{\frac{2}{3},\frac{4}{7},\mbox{\ensuremath{\frac{5}{9}},\ensuremath{\frac{1}{2},\frac{4}{5}}}\}$, respectively. The number of secondary users $N=150$ with the social information sharing graph given in Figure \ref{fig:A-square-area}. We implement the imitative spectrum access mechanism, such that in each decision period a secondary user receives the delayed throughput information from other users and carries out the imitation based on the delayed information. Figure \ref{fig:Delay} shows the results with the information exchange delay $D=0, 50, 80$, and $100$ decision periods, respectively. We observe that the proposed imitative spectrum access mechanism is quite robust to the information exchange delay. When the delay is not very large (e.g., $D\leq 50$ periods), the mechanism can still converge to the same imitation equilibrium as the case without delay. When the delay $D$ is too large (e.g., $D\geq80$ periods), the mechanism fails to converge to  the imitation equilibrium, since the throughput information is completely out-dated.

\begin{figure}[tt]
\centering
\includegraphics[scale=0.45]{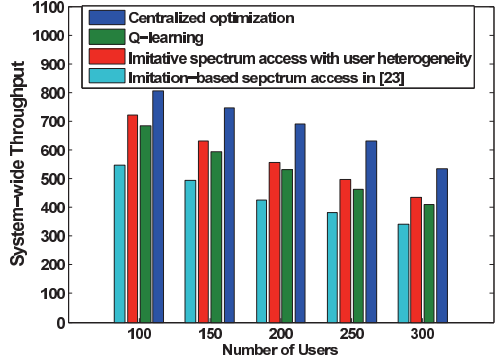}
\caption{\label{fig:system} Comparison of system-wide throughput of different solutions.}
\end{figure}

\begin{figure}[tt]
\centering
\includegraphics[scale=0.45]{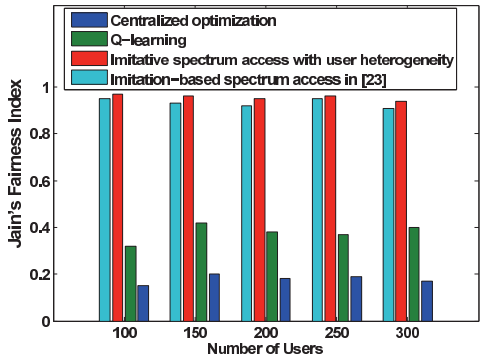}
\caption{\label{fig:fair}Comparison of fairness of the solutions of different solutions.}
\end{figure}

\subsection{\label{Comparison}Performance Comparison}
We now compare the proposed imitative spectrum access mechanism with the imitation-based spectrum access mechanism in \cite{key-IM6}.  Notice that the mechanism in \cite{key-IM6} requires the global network information including the channel characteristics and other users' channel selections to compute user's throughput. When users are homogenous (i.e., different users achieve the same data rate on the same channel), both mechanisms can converge to the imitation equilibrium and hence achieve the same performance. 

We now focus on performance comparisons for the more general and practical case that users are heterogenous. As the benchmark, we also implement the centralized optimal solution that maximizes the system-wide throughput (i.e., $\max \sum_{n=1}^{N} U_n$) and the decentralized spectrum access solution by Q-learning mechanism proposed in \cite{Q_learning}. Similarly to the setting in Section \ref{SG}, we consider $N =100,150,...,300$ randomly scattered users, respectively. The mean data rate of user $n$ on channel $m$ is $B_{m}^{n}=R$, where $R$ is a random value drawn from the uniform distribution over $(0,200)$. For each fixed user number $N$, we average over $50$ runs.

We first show the system-wide throughput achieved by  different mechanisms in Figure \ref{fig:system}. We see that the system-wide throughput of all the solutions decreases as the number of users increases. As the user population increases, the contention among users becomes more severe, which leads to more spectrum access collisions. We observe that the proposed imitative spectrum access mechanism with user heterogeneity achieves up-to $32\%$ performance improvement over the imitation-based spectrum access mechanism in \cite{key-IM6}. This is because that the mechanism in \cite{key-IM6} carries out imitation based on other user's throughput information directly, which ignores the fact that users are heterogeneous. While our mechanism takes user heterogeneity into account and carries out imitation based on the channel contention level. Compared with Q-learning mechanism, the imitative spectrum access mechanism can achieve better performance, with a performance gain of around $5\%$. Moreover, the performance loss of the imitative spectrum access mechanism with respect to the centralized optimal solution is at most $20\%$ in all cases. This demonstrates the efficiency of the imitative spectrum access mechanism with user heterogeneity.

We then compare the fairness achieved by  different mechanisms in Figure \ref{fig:fair}. We adopt the widely-used Jain's fairness index \cite{jain1984quantitative} to measure the fairness.  A larger index $J$ represents a more fair channel allocation, with the best case $J=1$. Figure \ref{fig:fair} shows that the centralized optimal solution is poor in terms of fairness (with the highest index value $J=0.2$). This reason is that the centralized optimal solution would allocate the best channels to a small fraction of users only (to avoid congestion) and most users will share those channels of low data rates. Our imitative spectrum access is much more fair and achieves up to $530\%$ and $300\%$ fairness improvement over the centralized optimization and Q-learning, respectively. This demonstrates that the proposed imitation-based mechanism can provide good fairness across users.

%% file: Conclusion.tex
\section{\label{sec:Conclusion}Conclusion}

In this paper, we design a distributed spectrum access mechanism with
incomplete network information based
on social imitations. We show that the imitative spectrum access mechanism
can converge to an imitation equilibrium on different information sharing graphs. When the information sharing graph is connected and users are homogeneous, the imitation equilibrium corresponds to a fair channel allocation such that all the users achieve the same throughput. We also extend the imitative spectrum access
mechanism to the case that users are heterogeneous.  Numerical results demonstrate that the proposed imitation-based mechanism can achieve efficient spectrum utilization and meanwhile provide good fairness across secondary users.

To get useful initial insights of imitation for the distributed spectrum access mechanism design  with incomplete network information, we have assumed that all the users interfere with each other in the spectrum sharing network. We plan to further extend our study to the case with spatial reuse. How to design an efficient imitation based spectrum
access mechanism where each user only interferes with a subset of users in the spectrum sharing network is very challenging.

%% file: Appendix.tex
\section{Proofs} \label{appendixA}
\subsection{\label{proof2} Proof of Theorem \ref{thm1}}
From (\ref{eq:DD6}) and (\ref{eq:DD7}), we have\begin{align*}
& \dot{X}_{m}^{k}(t) \\
= & X_{m}^{k}(t+1)-X_{m}^{k}(t)= \sum_{i=1}^{M}X_{i}^{k}(t)P_{i,k}^{m}(\boldsymbol{X}(t))-X_{m}^{k}(t) \\
= & \sum_{i=1}^{M}X_{i}^{k}(t)P_{i,k}^{m}(\boldsymbol{X}(t))-X_{m}^{k}(t)\sum_{i=1}^{M}P_{m,k}^{i}(\boldsymbol{X}(t))\\
= & \sum_{i=1}^{M}X_{i}^{k}(t)\sum_{k'\in\mathcal{C}_{k}}\frac{z_{k'}}{\sum_{l\in\mathcal{C}_{k}}z_{l}}X_{m}^{k'}(t)Q(U(m,\boldsymbol{X}(t))-U(i,\boldsymbol{X}(t)))\\
 &  -X_{m}^{k}(t)\sum_{i=1}^{M}\sum_{k'\in\mathcal{C}_{k}}\frac{z_{k'}}{\sum_{l\in\mathcal{C}_{k}}z_{l}}X_{i}^{k'}(t)Q(U(i,\boldsymbol{X}(t))-U(m,\boldsymbol{X}(t))).\end{align*}
which completes the proof. \qed

\subsection{\label{proof1}Proof of Theorem \ref{thm:For-the-imitative-1}}
To proceed, we first define the following function\begin{equation}
V(\boldsymbol{X}(t))=-\sum_{m=1}^{M}\int_{-\infty}^{\sum_{k=1}^{K}z_{k}X_{m}^{k}(t)}\theta_{m}B_{m}g(s)ds.\label{eq:KK}\end{equation}
We then consider the variation of $V(\boldsymbol{X}(t))$ along the
evolution trajectory of deterministic population state $\{\boldsymbol{X}(t)\}$,
i.e., differentiating $V(\boldsymbol{X}(t))$ with respective to time
$t$, \begin{align}
& \frac{dV(\boldsymbol{X}(t))}{dt} = -\sum_{k=1}^{K}\sum_{m=1}^{M}\frac{dV(\boldsymbol{X}(t))}{dX_{m}^{k}(t)}\frac{dX_{m}^{k}(t)}{dt} \nonumber \\
= & -\sum_{k=1}^{K}\sum_{m=1}^{M}z_{k}U(m,\boldsymbol{X}(t))\sum_{i=1}^{M}X_{i}^{k}(t) \nonumber \\
& \times \sum_{h\in\mathcal{C}_{k}}\frac{z_{h}}{\sum_{k'\in\mathcal{C}_{k}}z_{k'}}X_{m}^{h}(t)Q(U(m,\boldsymbol{X}(t))-U(i,\boldsymbol{X}(t))) \nonumber \\
 &   +\sum_{k=1}^{K}\sum_{m=1}^{M}z_{k}U(m,\boldsymbol{X}(t))X_{m}^{k}(t)\nonumber \\
 & \times \sum_{i=1}^{M}  \sum_{h\in\mathcal{C}_{k}}\frac{z_{h}}{\sum_{k'\in\mathcal{C}_{k}}z_{k'}}X_{i}^{h}(t)Q(U(i,\boldsymbol{X}(t))-U(m,\boldsymbol{X}(t))) \nonumber \\
 = & -\sum_{k=1}^{K}\frac{1}{\sum_{k'\in\mathcal{C}_{k}}z_{k'}}\sum_{h\in\mathcal{C}_{k}}\sum_{m=1}^{M}\sum_{i=1}^{M}z_{k}z_{h}X_{m}^{k}(t)X_{i}^{h}(t) \nonumber \\
 & \times \left(U(m,\boldsymbol{X}(t))-U(i,\boldsymbol{X}(t))\right)\left(Q(U(m,\boldsymbol{X}(t))-U(i,\boldsymbol{X}(t))) \right. \nonumber \\
 & \left. -Q(U(i,\boldsymbol{X}(t))-U(m,\boldsymbol{X}(t)))\right). \label{eq:C2}\end{align}

Since $f(\omega)$ is a probability density function satisfying $f(\omega)>0$, for all $\omega\in(\underline{\omega},\overline{\omega})$
and $\int_{\underline{\omega}}^{\overline{\omega}}f(\omega)d\omega=1$,
it follows from (\ref{eq:DD5}) that $q(\varpi)>0$, for all $\varpi\in(\underline{\omega}-\overline{\omega},\overline{\omega}-\underline{\omega})$
and $q(\varpi)=0$, for all $\varpi\notin(\underline{\omega}-\overline{\omega},\overline{\omega}-\underline{\omega})$.
Hence the cumulated probability function $Q(\varpi)=\int_{-\infty}^{\varpi}q(s)ds$
is strictly increasing for any $\varpi\in(\underline{\omega}-\overline{\omega},\overline{\omega}-\underline{\omega})$,
and further $Q(\varpi)=0$, for all $\varpi\in(-\infty,\underline{\omega}-\overline{\omega})$
and $Q(\varpi)=1$, for all $\varpi\in(\overline{\omega}-\underline{\omega},+\infty)$. This implies that if $U(m,\boldsymbol{X}(t))\neq U(i,\boldsymbol{X}(t))$,
\begin{align}
 & \left(U(m,\boldsymbol{X}(t))-U(i,\boldsymbol{X}(t))\right)\left(Q(U(m,\boldsymbol{X}(t))-U(i,\boldsymbol{X}(t))\right. \nonumber \\
 & \left. -Q(U(i,\boldsymbol{X}(t))-U(m,\boldsymbol{X}(t))\right)>0,\label{eq:C3-1}\end{align}
and if $U(m,\boldsymbol{X}(t))=U(i,\boldsymbol{X}(t))$, \begin{align}
& \left(U(m,\boldsymbol{X}(t))-U(i,\boldsymbol{X}(t))\right)\left(Q(U(m,\boldsymbol{X}(t))-U(i,\boldsymbol{X}(t))\right. \nonumber \\
& \left. -Q(U(i,\boldsymbol{X}(t))-U(m,\boldsymbol{X}(t))\right)=0.\label{eq:C4-1}\end{align}
From (\ref{eq:C2}), (\ref{eq:C3-1}) and (\ref{eq:C4-1}), we have $\frac{dV(\boldsymbol{X}(t))}{dt}\leq0$.
Hence $V(\boldsymbol{X}(t))$ is non-increasing along the trajectory
of the evolution dynamics. According to the Lasalle's
principle  \cite{key-IM4}, the evolution dynamics of deterministic population
state $\{\boldsymbol{X}(t)\}$ must asymptotically converge to a limit
point $\boldsymbol{X}^{*}$ such that $\frac{dV(\boldsymbol{X}^{*})}{dt}=0,$
i.e., $\forall m,i\in\mathcal{M},h\in \mathcal{C}_{k},k\in\mathcal{K}$\begin{align}
 & X_{m}^{k*}X_{i}^{h*}\left(U(m,\boldsymbol{X}^{*})-U(i,\boldsymbol{X}^{*})\right)\left(Q(U(m,\boldsymbol{X}^{*})-U(i,\boldsymbol{X}^{*})\right. \nonumber \\
 & \left. -Q(U(i,\boldsymbol{X}^{*})-U(m,\boldsymbol{X}^{*})\right)=0.\label{eq:C6}\end{align}
According to Lemma \ref{lem:The-set-of}, the set of users that share information with
user $n$ is the same as the set of users in user $n$'s cluster and the clusters that communicate
with user $n$'s cluster, i.e., $\mathcal{N}_{n}=\cup_{k'\in\mathcal{C}_{k(n)}}\mathcal{N}(k')$.
Thus we must have $U(m,\boldsymbol{X}^{*})=U(i,\boldsymbol{X}^{*}),\forall m,i\in\Delta_{n}(\boldsymbol{X}^{*}),\forall n\in\mathcal{N},$
where $\Delta_{n}(\boldsymbol{X}^{*})=\{m\in\mathcal{M}:\exists a_{i}^{*}=m,\forall i\in\mathcal{N}_{n}\}$.
This corresponds to the imitative equilibrium. \qed

\subsection{\label{proof22}Proof of Corollary \ref{cor222}}
Since the information sharing graph is connected, for any two different
users $n$ and $n'$, there must exists a path $(n_{1}=n,n_{2},...,n_{L}=n')$
satisfying that $n_{l+1}\in\mathcal{N}_{n_{l}},\forall1\leq l\leq L-1$.
According to Theorem \ref{thm:For-the-imitative-1}, we have $U(m,\boldsymbol{X}^{*})=U(i,\boldsymbol{X}^{*}),\forall m,i\in\Delta_{n}(\boldsymbol{X}^{*}),\forall n\in\mathcal{N}.$ Since $a_{n_{l+1}}^{*}\in\Delta_{n_{l}}(\boldsymbol{X}^{*}),\forall1\leq l\leq L-1$, it implies that $U(a_{n_{1}}^{*},\boldsymbol{X}^{*})=U(a_{n_{2}}^{*},\boldsymbol{X}^{*})=...=U(a_{n_{L}}^{*},\boldsymbol{X}^{*}).$ \qed

\subsection{\label{proof33}Proof of Corollary \ref{cor333}}
According to Cauchy-Schwarz inequality, we know that
$\left(\sum_{n=1}^{N}U(a_{n}^{*},\boldsymbol{X}^{*})\right)^{2}\leq N\sum_{n=1}^{N}U(a_{n}^{*},\boldsymbol{X}^{*})^{2}$
and $\left(\sum_{n=1}^{N}U(a_{n}^{*},\boldsymbol{X}^{*})\right)^{2}=N\sum_{n=1}^{N}U(a_{n}^{*},\boldsymbol{X}^{*})^{2}$
if and only if $U(a_{n}^{*},\boldsymbol{X}^{*})=U(a_{m}^{*},\boldsymbol{X}^{*})$, for any $n,m=1,...,N$.
It then follows that Jain's fairness index $J$ is maximized at the imitation
equilibrium, since the condition $U(a_{n}^{*},\boldsymbol{X}^{*})=U(a_{m}^{*},\boldsymbol{X}^{*})$
holds according to Corollary \ref{cor222}. \qed

\subsection{\label{proof44}Proof of Theorem \ref{PoI}}
First of all, according to Corollary \ref{cor222}, we know that all the users
at the imitation equilibrium $\boldsymbol{X}^{*}$achieve the same
throughput. Since all the channels are homogenous, the number of users
on each of $Z$ utilized channels is the same, $\frac{N}{Z}$. It
then follows that the system-wide throughput at the imitation equilibrium
is $\sum_{n=1}^{N}U(a_{n}^{*},\boldsymbol{X}^{*})=NB\theta g(\frac{N}{Z}).$
On the other hand, for the centralized optimal solution, since $kg(k)\leq1$
for any $k=1,2,...,N$, we know that $\max_{\boldsymbol{X}}\sum_{n=1}^{N}U(a_{n},\boldsymbol{X})=\max_{(k_{1},..,k_{M})}\sum_{m=1}^{M}k_{m}B\theta g(k_{m})\leq MB\theta$. Thus, we can conclude that the PoI is at least $\frac{Ng(\frac{N}{Z})}{M}$. \qed